\documentclass[12pt]{article}
\newcommand{\lappr}{\hspace{0.3em}\raisebox{-0.6ex}{$\stackrel{<}{\sim}$}\hspace{0.3em}} 
\newcommand{\gappr}{\hspace{0.3em}\raisebox{-0.6ex}{$\stackrel{>}{\sim}$}\hspace{0.3em}} 
\begin{document}
\title{Energy Loss of Ultrahigh Energy Protons in Strong Magnetic Fields}
\author{G.~Domokos and S.~Kovesi-Domokos \\[1mm]
The Henry A.~Rowland Department of Physics and Astronomy\\
The Johns Hopkins University\\
Baltimore, MD 21218\thanks{e-mail: skd@haar.pha.jhu.edu}}
\maketitle
\begin{abstract}
Ultrahigh energy protons in magnetic fields produce pions and thus lose 
energy. The mean free path of such a process is worked out for Gaussian
random fields. Two cases are considered: an isotropic and a cylindrically
symmetric distribution. The energy loss is  proportional
to $E^{3}\langle B^{2}\rangle $; it becomes
significant for protons of energies $\gappr 10^{19}$eV and magnetic fields
$B\gappr 10^{9}$Gauss. For  energies and magnetic fields of this
magnitude, a proton
injected into the magnetic field loses a substantial fraction
of its initial energy due to pion production.
\end{abstract}
\begin{flushleft}
\end{flushleft}
\section{Introduction}
It is well known that the propagation of ultra high energy (UHE) protons in
the Universe is limited by the Greisen-Zatsepin-Kuzmin (GZK) mechanism,
\cite{greisen, zatsepin}. The propagating protons undergo inelastic 
scattering on the photons of the cosmic microwave background radiation 
(CMBR) and produce pions. On the average, the initial energy is shared
(roughly) equally by the nucleon and pion in the final state. The energy
at which the GZK mechanism becomes important can be crudely estimated
by saturating the inelastic cross section by the $\Delta$ resonance.
Given the fact that the average energy of the CMBR photons is around
$3\times 10^{-4}$eV, one gets that pion production becomes
significant around proton energies of the order of $10^{19}$eV. (In fact,
a similar simple estimate was used in Greisen's original paper.)

Here we discuss a mechanism for energy loss of UHE protons, hitherto 
apparently ignored, {\em viz\/.} by inelastic scattering on virtual photons.
This mechanism plays no significant role in limiting the
propagation of UHE protons in intergalactic space. However, it 
becomes significant when the propagation is considered in an environment
where strong magnetic fields exist, \mbox{{\em e.g.\/}} in jets emerging
from gamma ray bursters (GRB) or jets in active galactic nuclei (AGN). 

The ``average'' energetics of pion
production on virtual photons (typically, on an external magnetic field)
is very different from that of the GZK mechanism. In fact, if the size
of the external magnetic field is characterized by a length $L$, then
the typical momentum of the virtual photon is of the order of $1/L$.
Consequently, the  average invariant center of mass (CMS) energy
available for pion production is of the order,
\[ s \approx  m^{2} + 2E/L, \]
where $m$ stands for the nucleon mass and $E$ is the energy of the incident
proton in the rest frame of the local universe. Obviously, this energy is 
much less than the analogous quantity in the GZK process for {\em any}
macroscopic $L$. Nevertheless, for a sufficiently strong and spatially
confined magnetic field, one obtains an appreciable production rate, due to 
the fact that the Fourier spectrum of a confined field is rather slowly
decreasing with the wave number (typically, as a power). As a consequence, 
Fourier components 
with $\left|{\bf k}\right|\gg 1/L$ can play a significant  role.

This paper is organized as follows. In the next section, we obtain 
a general expression of the cross section for the process
$p+B \rightarrow X$, where $B$ stands for an external magnetic field.
We also discuss the approximations one can make in order to simplify the
calculation. The subsequent section,~\ref{sec:random} contains an evaluation
of the interaction rate for random magnetic fields: we believe that 
this serves as a first model for energy loss in the chaotic fields 
present in typical astrophysical environments. Two situations
are considered in detail: an isotropic and a cylindrically symmetric
probability distribution of the random field.

While an isotropic environment largely serves to illustrate
the physical features of the process in a simple context, it is also
potentially applicable to a situation in which the size of the 
magnetic field is substantially larger than the interaction mfp. 
The calculation of the interaction rate for a cylindrically symmetric 
environment is relevant for jets, for instance, emerging from an AGN or
GRB. The results are discussed in sec.~\ref{sec:discussion}.
\section{General expression of the cross section in an external 
magnetic field.} \label{sec:general}
The calculation described here is an elementary application of the
optical theorem. The amplitude of a proton interacting with an external
field and producing a final state $\left|X \rangle \right.$ is given by
\begin{equation}
T(p+B\rightarrow X)= \int d^{4}x \quad A_{\mu}(x)
\langle  X\left|j^{\mu}(x)\right| p\rangle
\label{eq:amplitude}
\end{equation}
Here, $A_{\mu}(x)$ stands for the vector potential of the external
field and $j_{\mu}(x)$ is the density of the electromagnetic current.
Squaring (\ref{eq:amplitude}) and summing over the final states 
$\left| X\rangle \right.$, one expresses the cross section in terms of the
current correlation function. This is textbook material, see for instance
\cite{leader}. We also assume that that the external magnetic field is
static. This assumption simplifies the calculation. From a physical point
of view, it is justifiable even if astrophysical objects of bulk Lorentz
factors of the order of a few hundred are considered: the protons we are
interested in have Lorentz factors which are ten or eleven orders of
magnitude larger.

On writing for the Fourier transform of the vector potential
\begin{equation}
A_{\mu}(q)= \delta\left( q_{0}\right) a_{\mu}\left( {\bf q}\right)
\label{eq:static}
\end{equation}
and using a gauge in which $A_{0}=0$, one gets:
\begin{equation}
\sigma = \frac{4\pi^{2}\alpha m}{E}\int d^{3}q \quad a^{*}_{i}({\bf q}) 
a_{k}({\bf q})W_{ik}
\label{eq:generalsigma}
\end{equation}
In equation~(\ref{eq:generalsigma}) $m$ and $E$ stand for the mass
and energy of the incident proton, respectively and $W_{ik}$ is the
spatial part of the standard polarization tensor:
\begin{equation}
W_{\mu \nu}= \frac{F_{1}}{m}
\left( -g_{\mu \nu} + \frac{q_{\mu}q_{\nu}}{q^{2}}\right)
+ \frac{F_{2}}{\nu}\left( p_{\mu} - q_{\mu}\frac{(pq)}{q^{2}}\right)
\left(p_{\nu} - q_{\nu}\frac{(pq)}{q^{2}}\right)
\label{eq:W}
\end{equation}
The notation is standard, $p$ and $q$ are the four momenta of the
incident proton and virtual photon, respectively, $\nu = (pq)/m$.

A further simplification is possible due to the fact that the protons 
we are interested in are extreme relativistic and the average value
of the momentum of the virtual photon is of the order of $1/L$. 
In order to motivate this simplification, we Lorentz transform 
to the rest frame of the proton. In that frame the components of the
four momentum $q$ and the field quantities are distinguished by a prime.
Components perpendicular to the direction of motion are denoted by 
capital letters; longitudinal components by a subscript $l$. Since we 
have $v\sim 1$, the transformation
formul\ae \hspace{0.5em}are:
\begin{equation}
q_{0}^{'} \sim \frac{1}{2}\exp (y) q_{l},\quad
q_{l}^{'} \sim \frac{1}{2} \exp (y) q_{l},\quad  q_{A}^{'} \sim q_{A}.
\end{equation}
\begin{eqnarray}
B_{l}^{'} = B_{l}, \quad E_{l}^{'} = E_{l} =0, \nonumber \\
B_{A}^{'}\sim \frac{1}{2}\exp (y) B_{A}, \quad
E_{A}^{'}\sim \frac{1}{2} \exp (y) \epsilon_{AB} B_{B}.
\end{eqnarray}
In the last two equations, $y$ stands for the rapidity.

As a consequence, apart from corrections of $O(\exp (-y))$,
\begin{equation}
q^{2}\sim 0, \quad {\bf B}\cdot {\bf E}\sim 0, \quad {\bf B}^{2} - {\bf E}^{2}\sim 0.
\end{equation}
In a reference frame comoving with the proton, the magnetic field appears
as a stream of (almost real) photons: consequently, the contribution of 
the structure function $F_{2}$ to the cross section is negligibly small.

One can then express the cross section on the external magnetic field
in terms of the photoproduction cross section, $\sigma_{\gamma}$, 
{\em viz\/.}
\begin{equation}
\sigma \sim \frac{1}{E} \int d^{3}q\quad a_{i}({\bf q}) a_{j}({\bf q})^{*}
\frac{\bf(p\cdot q)}{\bf q^{2}} \sigma_{\gamma} \left( \delta_{ij}{\bf q}^{2}
 - q_{i}q_{j}\right)
\label{eq:weizsacker}
\end{equation}
One readily recognizes that eq.~(\ref{eq:weizsacker}) is equivalent to
a Weizs\"{a}cker-Williams approximation to the cross section. Because of
the presence of a transverse projector, that expression is a manifestly
gauge invariant one. It is worth noticing that in the 
Weizs\"{a}cker-Williams approximation the expression of the cross 
section is independent of the mass of the projectile. Hence the same 
expression can be used to describe {\em e\/.g\/.} photon induced
reactions in a magnetic field.

Finally, one considers the evaluation of $\sigma_{\gamma}$. Due to the
fact that the photoabsorption cross section is to be evaluated
near the pion production threshold, to a good approximation one can
saturate it by the contribution of the $\Delta$ resonance.
A narrow resonance approximation is sufficiently accurate.
Hence we put
\begin{equation}
\sigma_{\gamma} \approx \sigma_{0} \delta \left( s - m_{\Delta}^{2}\right),
\label{eq:narrow}
\end{equation}
where the dimensionless quantity, $\sigma_{0}$ is the
integral of the pion photoproduction cross section across the resonance,
\[ \sigma_{0} = \int_{(res)} \sigma (s) ds. \]
Using a standard
invariant Breit-Wigner fit and  the data available,~\cite{pdg}
one gets $\sigma_{0}\approx 0.3$.
\section{Random magnetic fields}
\label{sec:random}
We model the chaotic magnetic fields present in the  astrophysical
environments of interest by means of a Gaussian random field of zero mean.
The central object in the theory of random fields is the generating
functional of the correlation functions. In the case of a Gaussian field, 
only the second cumulant is different from zero. We write the generating
functional as follows~\cite{gaussian}.
\begin{equation}
Z[j] = \int {\cal D}{\bf a}\quad  
\exp \left[ -S +i \int d^{3}k j_{r}\left({\bf k}\right)
a_{r}\left( {\bf k}\right) \right]
\label{eq:generating}
\end{equation}
In eq.~(\ref{eq:generating}) ${\bf j}$ stands for an external source, 
${\bf a}$ is the Fourier transform of the vector potential, {\em cf\/.}
eq.~(\ref{eq:static}). The functional $S$ is a generalized entropy; for
a Gaussian field it is a (gauge invariant) quadratic functional of ${\bf a}$.
We write $S$ as follows.
\begin{equation}
S= \int d^{3}k \quad a^{*}_{i}({\bf k}) a_{j}({\bf k})
\left( \delta_{ij} - \frac{k_{i} k_{j}}{{\bf k}^{2}}\right)
\frac{{\bf k}^{2}}{4 \pi L^{3} \langle B^{2}\rangle}
\left( 1 + L^{2} k_{r}k_{s} u_{rs}\right)^{2}.
\label{eq:entropy}
\end{equation}
In eq.~(\ref{eq:entropy}), $L$ stands for the root mean square correlation
length (the average is taken over directions). The tensor $u_{rs}$
characterizes the directional distribution of the probability density.
For a general, arbitrarily anisotropic distribution, $u_{rs}$ has 6
independent components: {\em e\/.g\/.} the three, mutually
orthogonal,  principal correlations  and the three angles
describing the orientation of the principal  correlations.
In what follows, however, we consider environmernts of high symmetry;
consequently, fewer parameters are sufficient. The factor
$\left( 1+ L^{2} k_{r}u_{rs}k_{s}\right)^{2}$ ensures an exponential
decrease of the correlation function with distance, 
{\em cf\/.}~ref.~\cite{gaussian}. 

In considering particle production in a random field, one has to 
replace  factors such as  $a^{*}_{i}a_{j}$ in eq.~(\ref{eq:generalsigma})
and subsequent ones by their expectation values in the ensemble defined
by eqs.~(\ref{eq:generating}) and (\ref{eq:entropy}).

We now consider two special cases of the ensembles in order to calculate
particle production cross sections.
\subsection{Isotropic ensemble.}
\label{subsec:isotropic}
This ensemble is characterized by the tensor $u_{ij}$ in 
eq.~(\ref{eq:entropy}) being the unit tensor, $u_{ij} = \delta_{ij}$.
One finds:
\begin{equation}
\langle a_{i}({\bf q})a_{j}({\bf q}^{'})^{*}\rangle = 
\delta^{3}\left({\bf q} - {\bf q}^{'}\right)
\left(\delta_{ij} - \frac{q_{i}q_{j}}{{\bf q}^{2}}\right)
\frac{4\pi \langle B^{2}\rangle L^{3}}{{\bf q}^{2}}
\left( 1 + L^{2}{\bf q}^{2}\right)^{-2}
\label{eq:isotropic}
\end{equation}
In the expression of the cross section, however, one finds
a factor $\langle a_{i}({\bf  q}) a^{*}({\bf q})\rangle$, which is
infinite, see  the last equation. 
This is due to the
fact that we idealized a region of non vanishing magnetic field
by one of infinite extent albeit of an exponentially decreasing
correlation function. In order to correct for the inconsistency
caused by this idealization, we replace the delta function
of vanishing argument by a quantity proportional to the 
volume,
{\em viz\/.}
\[
\delta^{3}(0) \longrightarrow \frac{1}{8\pi^{3}} V.
\]
(This is a consistent procedure provided  the density of levels
can be approximated by the Rayleigh - Jeans formula, as done here. 
In the problem under 
consideration, the conditions for the validity of that approximation 
are satisfied.)
We take $V$ to be the correlation volume; thus, in the isotropic
case, $V= 4\pi L^{3}/3$; clearly, different geometries give rise
to different expressions of the correlation volume. The important 
fact is, however that the incident flux is $\propto 1/V$. Thus
the reciprocal mfp is independent of the choice of the volume.

With this and using eq.~(\ref{eq:narrow}) the cross section can be
evaluated in terms of elementary functions. Quoting directly the
inverse of the absorption mfp which is the relevant quantity for
the applications, one finds:
\begin{equation}
\frac{1}{\lambda_{a}} = \frac{L}{\pi}\langle B^{2}\rangle 
\frac{\sigma_{0}}{m_{\Delta}^{2} - m^{2}}f(w).
\label{absorptionmfp}
\end{equation}
Here $f$ is a function of the dimensionless variable, 
$w = \left( m_{\Delta}^{2} - m^{2}\right)L/2E$. Its explicit
form is:
\begin{equation}
f(w) = w^{2}\left[ \ln \left( \frac{1+w^{2}}{w^{2}}\right)
     + \frac{1}{1+w^{2}} \right]
\label{eq:f}
\end{equation}
However, it was pointed out in the Introduction that $1/L$ is a
{\em small} momentum. As a consequence, we only need eq.~(\ref{eq:f})
for large values of $w$. In that case, eq.~(\ref{eq:f}) simplifies to:
\[ f(w) \sim \frac{1}{2w^{2}} \qquad (w \gg 1). \]
Hence, the expression of $\lambda_{a}$ becomes:
\begin{equation}
\frac{1}{\lambda_{a}} \sim \frac{2 \sigma_{0}}{L \pi}\langle B^{2}\rangle
\frac{E^{2}}{\left(m_{\Delta}^{2} - m^{2}\right)^{3}}
\label{eq:asympt}
\end{equation}
\subsection{Cylindrically symmetric ensemble}
\label{subsec:cylindrical}
This geometry is a more realistic one. In particular, an astrophysical
jet as emerging, for instance, from a GRB or an AGN can be approximated by
a cylindrical geometry at the early stages of expansion. (At early
stages, the lateral expansion is negligibly small compared to the
longitudinal one.) Approximating the jet by one of cylindrical 
geometry means that the lateral expansion is neglected altogether.
It has been known for a long time that this is an acceptable approximation
in the initial stages of expansion of a relativistic fluid \cite{landau}

In this case, the tensor $u_{ij}$ in eq.~(\ref{eq:entropy}) effectively
depends on one parameter only. It is convenient to introduce the
longitudinal and transverse correlation lengths with respect to the axis
of the cylinder and an {\em anisotropy parameter}, $\alpha$,
such that
\[ L_{T}^{2} = \alpha L^{2}, \quad L_{L}^{2} = L^{2}(1-\alpha). \]
In practice, $\alpha \ll 1$, say $\alpha \approx 0.1$ 
or so\footnote{E. Vishniac, private communication.}.
Using this parametrization, we have:
\begin{equation}
\left( 1 + L^{2}q_{i}u_{ij}q_{j}\right) =
\left( 1 + \left( \alpha {\bf q_{T}}^{2} + (1-\alpha )q_{L}^{2}\right) \right)
\label{eq:cylinder}
\end{equation}
In eq.~(\ref{eq:cylinder}), ${\bf q_{T}}$ and $q_{l}$ stand for the 
momentum components perpendicular and parallel to the axis  of
the cylinder, respectively.

In the case of such a  geometry, the integral occurring in
eq.~(\ref{eq:weizsacker}) cannot be calculated in a closed form.
In essence, this is due to the fact that the expression 
of the absorption cross section now contains two directions: that
of the incident proton and the axis of the cylinder. However,
instead of resorting to a numerical evaluation, we observe
that the variable $w$ is large and the absorption can only be
significant if the angle between the incident proton and the axis of the
cylinder is not too large: efficient absorption requires a coherent 
magnetic field. 

These simplifications allow a calculation of the absorption mfp in
a closed form. A somewhat tedious, but elementary calculation leads
to the result:
\begin{equation}
\frac{1}{\lambda_{a}}\sim \frac{4\sigma_{0}}{\pi L} F(\Theta)
\frac{E^{2}\langle B^{2}\rangle }{\left( m_{\Delta}^{2} - m^{2}\right)^{3}}
\label{eq:cylindrical}
\end{equation}
The factor $F(\Theta )$ is given by the expression:

\begin{equation}
F(\Theta ) \approx \left( \cos \Theta \right)^{3}
 \ln \left( \frac{ (\cos \Theta )^{2} }{\alpha} -1\right)
\label{eq:angular}
\end{equation}
In eq.~(\ref{eq:angular}) $\Theta$ stands for the angle between the
incident proton and the axis of the cylinder. Obviously, this expression
holds only if the angle $\Theta$ is small. From the physical point of view,
however, this is not a serious limitation: due to the presence of the
factor $\propto \left( \cos \Theta \right)^{3}$, the mean free path
becomes very large unless the angle of incidence with respect to
the axis of the cylinder is small.
\section{Discussion}
\label{sec:discussion}
Our approach has the advantage that it does not depend on the
details of the production process, since it is based on the use of the
optical theorem. However, its limitation is that the absorption
cross section is obtained to lowest order in the fine structure constant.
This poses no problem as long as 
$\sqrt{\langle B^{2}\rangle} \lappr m^{2}/e = B_{\mbox{crit}}$, where $m$ 
is the mass of 
a charged particle involved in the process. For electrons and
light quarks (u,d), the value of $B_{\mbox{crit}}$ is around
$10^{14}$Gauss. In magnetic fields of this order of magnitude, radiative
corrections and pair production become important. To our knowledge,
no results are available for such field strengths. Existing
calculations, such as Erber's,~\cite{erber} assume a homogeneous 
magnetic field. Calculations of this type can be used to estimate
energy losses as long as the magnetic fields are approximately 
homogeneous on the scale of the Larmor radius of the propagating 
charged particle. For realistic circumstances, however, this is 
hardly the case. Thus,
the question about the energy loss of charged particles in
astrophysically important magnetic field approaching the critical
value of the field, is still an open one. Our formulae, however, are 
expected to give at least a qualitative insight into the question of
absorption even for  near-critical fields.

Our results show that the circumstances neeeded
for the applicability of eq.~(\ref{eq:asympt}) are hardly met: one needs
magnetic fields with a coherence length  substantially in excess of
the Larmor radius at high energies. Nevertheless, that equation is  an
instructive one: due to its simplicity, the general features of the 
absorption cross section is easily understood.

From the physical point of view, eq.~(\ref{eq:cylindrical}) is 
more interesting. In order to assess the importance of the process
discussed it is worth converting eq.~(\ref{eq:cylindrical}) into  a form
permitting numerical estimates. The value of $\sigma_{0}$ has been 
quoted before; the rest of the numbers is also taken from 
ref.~\cite{pdg}. One obtains:
\begin{equation}
\frac{1}{\lambda_{a}}\approx \frac{5.5}{L} F\left( \Theta \right)
\left(\frac{E}{10^{20}\mbox{eV}}\right)^{2}
\frac{\langle B^{2}\rangle }{\left( 10^{9}\mbox{Gauss}\right)^{2}}
\label{eq:numerical}
\end{equation}
(We used the usual
conversion factor between the natural and conventional units of the
magnetic field, {\em viz\/.} \mbox{$B/1(\mbox{MeV})^{2} = 1.9\times 10^{-14}
B/1\mbox{Gauss}$.})

We find that a paraxially propagating proton of energy $\simeq 10^{20}$eV 
traversing a -- relatively 
modest -- magnetic field of $10^{9}$Gauss has an absorption mfp
about $(1/5)^{\mbox{th}}$ the  size of the magnetic field.

In a collision at the relevant energies, {\em on the average\/}
the nucleon and the produced pion in the final state share
the incident energy equally. As a consequence, using the 
continuous energy loss approximation, the energy loss
per unit path length is:
\begin{equation}
\frac{dE}{dx} = - \epsilon \frac{E}{\lambda}
\label{eq:energyloss}
\end{equation}
In the last equation, $\epsilon $ stands for the fractional
energy loss of a nucleon (in the laboratory system) due
to pion production. Assuming as we do throughout this paper that
the pion production cross section is dominated by the $\Delta $
resonance, one gets,
\[ \epsilon \approx \frac{m_{\pi}^{2}}{2m_{\Delta} m} \]
Due to the fact that $1/\lambda \propto E^{2}$, the energy loss
per unit path length grows as $E^{3}$. 

In fact, by inserting the expression for the mfp given by 
eq.~(\ref{eq:numerical}) into eq.~(\ref{eq:energyloss}),
the equation for the energy loss is readily integrated.
We exhibit the result for the energy loss of paraxial protons
($F(\Theta)\approx 1$) over one correlation length, $L$.
We get:
\begin{equation}
\frac{E(x)}{E(0)}\approx \left[ 1+ 0.1 \left( \frac{E}{10^{20}\mbox{GeV}}
\right)^{2}\left(\frac{\langle B^{2}\rangle}{\left( 10^{9}\mbox{Gauss}
\right)^{2}}\right)\frac{x}{L}  \right]^{-1/2}
\label{eq:fractionalenergyloss}
\end{equation}
We used the mass values listed in ref.~\cite{pdg} in order to arrive at
eq.~(\ref{eq:fractionalenergyloss}).

We conclude that the mechanism described in this paper
appears to be a major obstacle to accelerating  protons up to  energies of
the order of $10^{19}$eV or more by a conventional Fermi acceleration
mechanism. One notices for instance  that a proton of $E= 10^{20}$eV
injected into a field of $\sqrt{\langle B^{2}\rangle } = 10^{10}$Gauss
loses about 70\% of its initial energy over a correlation length.

This adds to the puzzle of the highest energy cosmic rays: it is known 
that particles of energy about $10^{20}$eV 
arrive to the Earth and they give rise to extensive air showers. 
At the same time, it appears to be increasingly difficult to
find an efficient mechanism for producing them at the usually
suspected sites, for instance in active galactic nuclei  or gamma 
ray bursters. 


\begin{thebibliography}{99}
\bibitem{greisen} K.~Greisen, Phys.~Rev.~Letters {\bf 16}, 748 (1966).
\bibitem{zatsepin} G.T.~Zatsepin and V.A.~Kuzmin, JETP-Letters~{\bf 4},
 78~(1966).
\bibitem{leader} E.~Leader and E.~Predazzi, An introduction to
gauge theories and modern particle physics; Vol.~1, Ch.~15.
Cambridge University Press, 1996.
\bibitem{pdg} C.~Caso~{\em et al\/.}, Eur.~Phys.~J. {\bf C3},~1~(1998);
  http://pdg.lbl.gov
\bibitem{gaussian} G.~Domokos and S.~Kovesi-Domokos, Heavy Ion Physics
{\bf 5}, 179 (1997). Festschrift for the $70^{th}$ birthday of
George Marx. See also, G.~Domokos and S.~Kovesi-Domokos, p. 191 in Beyond
the Standard Model V, edited by G.~Eigen, P.~Osland and B.~Stugu.
AIP conference proceedings 145, American Institute of Physics,
Woodbury, NY, 1997.
\bibitem{landau} See, for instance, S.Z.~Belen'kij and L.D.~Landau,
Supp.~Nuovo.~Cim. {\bf III},~15~(1956) and references quoted there. 
\bibitem{erber} T.~Erber, Rev.~Mod.~Phys.~{\bf 38},~626~(1996).
\end{thebibliography}
\end{document}